\documentclass[11pt,a4paper]{article}
\pdfoutput=1
\usepackage[nohyperref]{emnlp-ijcnlp-2019}
\usepackage{times}
\usepackage{latexsym}
\usepackage{graphicx}
\usepackage{booktabs}
\usepackage{url}
\usepackage{paralist}
\usepackage{tabularx}
\aclfinalcopy
\RequirePackage[normalem]{ulem} 
\RequirePackage{color}\definecolor{RED}{rgb}{1,0,0}\definecolor{BLUE}{rgb}{0,0,1} 
\providecommand{\DIFadd}[1]{{\protect\color{blue}\uwave{#1}}} 
\providecommand{\DIFdel}[1]{~}     
\providecommand{\DIFaddbegin}{} 
\providecommand{\DIFaddend}{} 
\providecommand{\DIFdelbegin}{} 
\providecommand{\DIFdelend}{} 

\usepackage{amsmath,amsfonts,bm}

\def\eqref#1{equation~\ref{#1}}

\def\1{\bm{1}}

\def\vx{{\bm{x}}}
\def\vy{{\bm{y}}}
\def\vz{{\bm{z}}}

\DeclareMathAlphabet{\mathsfit}{\encodingdefault}{\sfdefault}{m}{sl}
\SetMathAlphabet{\mathsfit}{bold}{\encodingdefault}{\sfdefault}{bx}{n}

\begin{document}

\title{Natural Adversarial Sentence Generation with \\ Gradient-based Perturbation}
\author{
Yu-Lun Hsieh$^{1,2}$, 
Minhao Cheng$^{3}$, 
Da-Cheng Juan$^{4}$, Wei Wei$^{4}$, \\
\textbf{Wen-Lian Hsu$^{1}$},
\textbf{Cho-Jui Hsieh$^{3,4}$} \\
$^1$Academia Sinica, Taiwan \\ 
$^2$National Chengchi University, Taiwan \\
$^3$University of California, Los Angeles, USA \\
$^4$Google Research, USA \\
\texttt{\small morphe@iis.sinica.edu.tw, mhcheng@cs.ucla.edu, x@dacheng.info, wewei@google.com,} \\ \texttt{\small hsu@iis.sinica.edu.tw, chohsieh@cs.ucla.edu}
}
\maketitle

\begin{abstract}
This work proposes a novel algorithm to generate natural language adversarial input for text classification models, in order to investigate the robustness of these models. It involves applying gradient-based perturbation on the sentence embeddings that are used as the features for the classifier, and learning a decoder for generation. 
We employ this method to a sentiment analysis model and verify its effectiveness in inducing incorrect predictions by the model.
We also conduct quantitative and qualitative analysis on these examples and demonstrate that our approach can generate more natural adversaries. 
In addition, it can be used to successfully perform black-box attacks, which involves attacking other existing models whose parameters are not known. 
On a public sentiment analysis API, the proposed method introduces a 20\% relative decrease in average accuracy and 74\% relative increase in absolute error. 
\end{abstract}

\section{Introduction}

Adversarial attacks on neural networks have drawn a significant amount of attention. 
The majority of these attacks are targeting computer vision models in which convolutional neural networks (CNN) are employed.
Due to the fact that input features 
to these models are continuous, we can apply  perturbation that is indistinguishable by humans.~\cite{szegedy2013intriguing}.
In contrast, generating adversarial samples has not been attempted for natural language processing (NLP) models until recently. 
The discrete nature of the input obfuscates the use of existing methods of generating adversarial examples~\cite{papernot2016}. 
Unlike image data, textual input consists of individual words that are represented by embeddings~\cite{Mikolov2013}, but we cannot directly apply perturbation to them and find another word.
Several approaches have been proposed to solve this problem, such as a Gumbel-Softmax~\cite{Kusner2016gumbel}, policy gradient, and Monte-Carlo search~\cite{yu2017seqgan}.
However, most existing methods 
try to minimize the number of word edits, which often lead to unnaturalness that is easily detected by humans~\cite{jia2017adversarial,zhao2018generating}.
How to create a generative model for natural adversarial examples 
remains an unsolved problem. 
Thus, we propose a novel approach that employs a sequence decoder to create  natural adversarial examples.

In addition, adversarial attacks can be categorized into two scenarios, i.e., white-box and black-box. 
The former has access to the parameters and structure of the model, and
the latter lacks such information~\cite{narodytska2016simple}.
Black-box attacks are more challenging in that they usually require a large number of queries to the model. 
However, they are more valuable towards evaluating machine learning models~\cite{papernot2017practical}.

This work aims to generate adversarial sentences that are natural as well as
capable of white-box and black-box attacks. 
We propose a novel algorithm that perturbs the sentence encoding by calculating gradients w.r.t.\ an alternative class, and utilize a well-trained decoder for sequence generation. 
We also experiment with a black-box attack against
an online sentiment classifier.
The main contributions of this work can be summarized as follows. 
\begin{enumerate}
\item Propose a novel method of generating natural adversarial examples using gradients.
\item Examine the objective quality of the examples through BLEU scores and perplexity. 
\item Conduct human evaluation of the quality of adversarial sentences produced by different methods.
\item Perform black-box attacks against a real world online sentiment classifier to validate the effectiveness of our approach.
\end{enumerate}

\section{Methodology}

\begin{figure}[tb]
\centering
\includegraphics[width=0.98\linewidth]{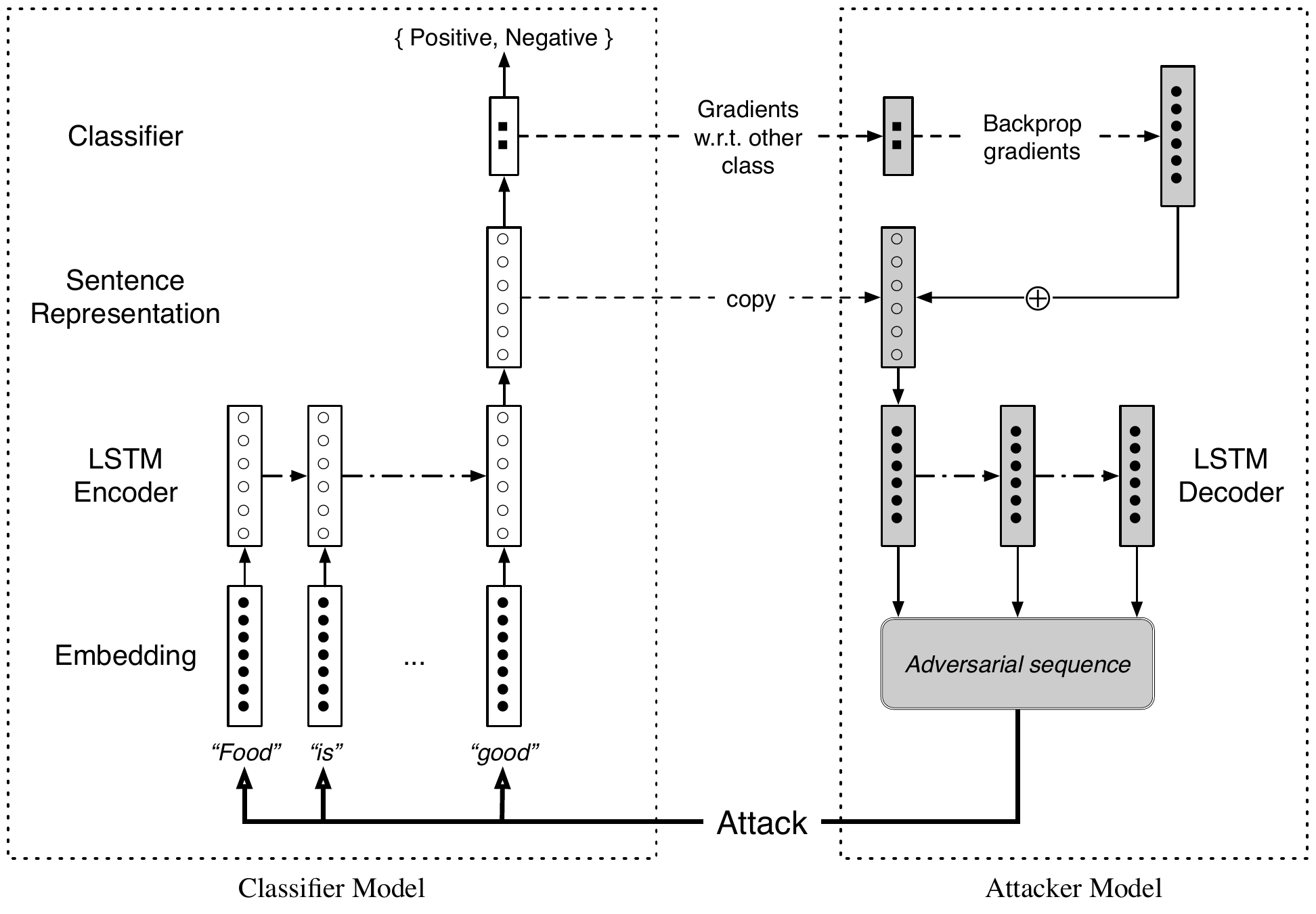}
\vspace{-5pt}
\caption{\label{fig:model_struct} Illustration of the proposed attack model. The classifier network consists of an LSTM encoder followed by a fully connected layer, and the attacker consists of an LSTM decoder.}
\end{figure}

At the outset, we train the target of our attack, which is an LSTM-based sequence classifier.
We assume this target model follows the encoder-based framework: given an input $\vx$, an encoder is used to generate the encoded input $\vz = \mathrm{Encoder}(\vx)$, and then the predicted label is computed by the classifier $\vy = h(\vz)$. 
Loss for the model is defined as the cross-entropy between the predicted class and true class labels. 

\subsection{Generating Adversarial Examples}
In our adversarial example generation algorithm, we first calculate gradients of the encoder output w.r.t.\ a target (opposite) label, and apply perturbation on the encoder output using  FGSM~\cite{goodfellow2014explaining} method before sending it to the decoder. 
The problem can be formulated as minimizing the loss function $L(\hat{\vy}, \vy_{adv})$, s.t.\ $\| \vz_{adv} - \vz \| \leq \epsilon$, where $\vz_{adv}$ is the adversarial encoded input, $\vz$ is the original encoded input, $\hat{\vy}=h(\vz_{adv})$ is the output of the classifier $h(\cdot)$, $\vy_{adv}$ is the adversarial target class.

To facilitate the white-box attack setting, a decoder $D(\cdot)$ is trained using the output of the encoder $\vz$ such that $D(\vz)=\vx'$.
Loss for the decoder is defined as the cross-entropy between the input sequence and decoded sequence. 
To generate adversarial examples, we
first pass the original input sentence into the encoder and obtain a sequence encoding as well as the classifier output.
Then, we calculate gradients of the classifier output w.r.t.\ a fake target and propagate back to the sentence encoding. 
Specifically, the cross entropy of the classifier output $\hat{\vy}$ against the fake target $\vy_{adv}$ is defined as
\begin{align}
&\mathcal{L}_{adv}(\hat{\vy}, \vy_{adv}) =&\\\nonumber
&-\big[\hat{\vy} \log \vy_{adv} + (1 - \hat{\vy})  \log (1 - \vy_{adv})\big]&
\label{eq:cross_ent_fake_target}
\end{align}
The fake target $\vy_{adv}$ is set to be the opposite of the original label. 
We then obtain the gradient of $\mathcal{L}_{adv}$ with respect to the encoded input $\vx$ as $\nabla_{\vz} \mathcal{L}_{adv}(h(\vz), \vy_{adv})$.
Finally, we apply FGSM as follows: 
\begin{equation}
\vz_{adv} = \vz - \epsilon \times \mathrm{sign}\left(\nabla_{\vz} \mathcal{L}_{adv}(h({\vz}), \vy_{adv})\right)
\label{eq:fgsm}
\end{equation}
Note that applying the gradient step with more iterations such as PGD-attack~\cite{madry2017towards} could potentially boost the performance, but we found one step of FGSM already leads to a good solution in the embedding space. 
The perturbed encoding $\vz_{adv}$ is sent to the decoder $D(\cdot)$ for generation of an adversarial sequence $\vx_{adv} = D(\vz_{adv})$.
We can then gradually perturb the sentence encoding by increasing  $\epsilon$.
In this way, we can modulate the perturbation on the sentence encoding space and observe its impact on the classifier as well as the generator.

\section{Experiments}


\subsection{Dataset}

We utilize the Yelp reviews polarity dataset released by~\citet{zhang2015character}
 in our experiments.
It contains
280,000 training samples and 19,000 test samples for each polarity.
At the outset, each review is tokenized using the spaCy English tokenizer. 
Then, we apply byte-pair encoding~\cite{sennrich2016neural} to further segment words into subword units. 
The vocabulary size is limited to approximately 8,000 subwords. 
Finally, each review is cropped to maximally 30 units. 

\subsection{Model Architecture}

The input sequence is converted to embeddings~\cite{Mikolov2013} with 512 dimensions.
We train a 1-layer LSTM encoder and decoder, both with hidden dimensions of 512. 
The encoder is regularized using the method and hyperparameters proposed by~\citet{zhao2017arae}. 
In addition, a 100-dimension fully connected network follows the encoder as the classifier. 
The classifier and decoder are trained for 20 epochs until it converges on 94\% classification accuracy on the training set and 86\% on the test set.

\subsection{Evaluation Metrics}
Objective evaluation involves gradually increasing the $\epsilon$ value and observe multiple metrics on the test set, namely, mis-classification rate of the classifier and average BLEU score~\cite{papineni2002bleu}.
Mis-classification rate denotes the percentage that the classifier produces an incorrect label, and BLEU scores evaluate the overlap of the original input and its corresponding adversarial example.
We also calculate a language model trained on the training set to estimate the quality of the generated examples.

On the other hand, subjective analysis include human evaluation of the readability and sentiment of sentences generated by both methods.
We recruit Amazon Mechanical Turkers to determine the naturalness and sentiment of the examples.
We further validate the effectiveness of these two methods on a free online sentiment analysis service by measuring the variation in accuracy and absolute error.

The compared method proposed by~\citet{papernot2016}
utilizes the Jacobian-based saliency map attack (denoted as JSMA).
It consists of greedily substituting each word in the input sequence with the most salient candidate.
The salient feature is produced by computing gradients of word features with regards to a target class.

\section{Results}

\begin{figure}[tb]
\centering
\includegraphics[width=.65\linewidth]{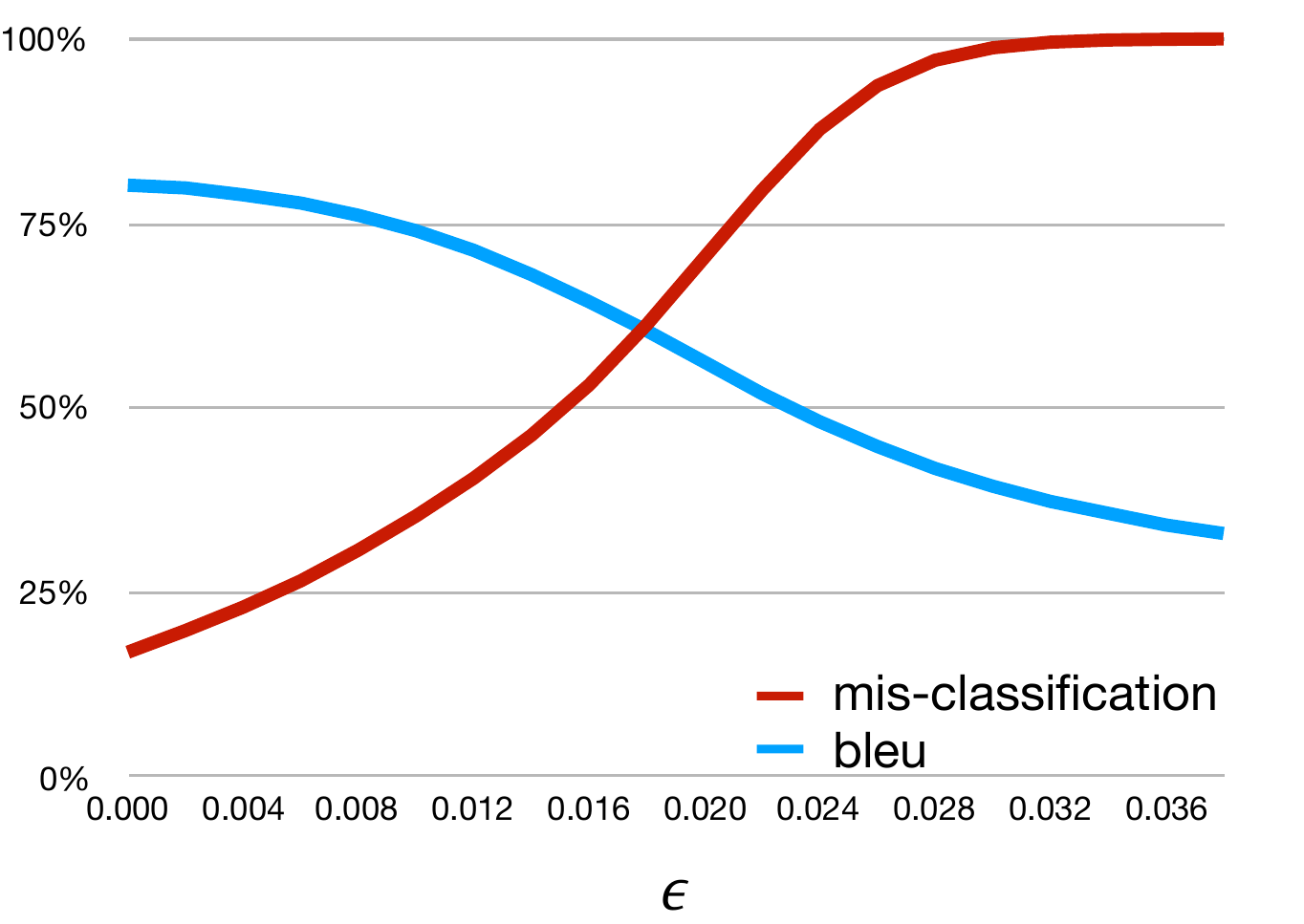}
\vspace{-5pt}
\caption{\label{fig:epsilon_variation}The effect of increasing the $\epsilon$ value on mis-classification rate and BLEU scores on the test set.}
\end{figure}

Quantitative analysis of the effect of adversarial examples on the classifier is visualized in Figure~\ref{fig:epsilon_variation}.
We observe that the mis-classification (error) rate positively relates to the epsilon value. 
However, the decrease in BLEU score is not drastic.
This indicates that our method can successfully fool the classifier with only a few modifications. 
Furthermore, we train a language model that obtains a log perplexity of 4.2 on the training set in order to evaluate the similarity of the generated sentences. 
The adversarial examples that our model generated using the test set obtain a log perplexity of 4.8, while that of the JSMA method is 6.1.
This again demonstrates that our method can produce sentences that are more similar to the distribution of the training data.

Next, we observe the results from human evaluation of the readability and `human accuracy' of judging the sentiment of the adversarial examples. 
Recall that a better adversarial example should have a higher readability and not affect human judgment of the sentiment, therefore obtain a greater human accuracy score.
Snippets of an original and generated sentence is listed in Table~\ref{tab:adv_ex_short}.
\begin{table}[h!]
\centering
\small
\setlength{\tabcolsep}{2pt}
\begin{tabular}{c|p{0.8\linewidth}}
\hline
{\bf Label} & {\bf Sentence} \\
\hline
P & I am currently trying to give this company another chance. I have had the same scheduling experience as others have written about. Wrote to them today\\
\hline
N & I am currently trying to give this company another \DIFaddbegin\DIFadd{review}\DIFaddend. I have had the same \DIFaddbegin\DIFadd{dental experience about others or written with a name. Thanks} \DIFaddend to them today\\
\hline
\end{tabular}
\vspace{-5pt}
\caption{Original and adversarial inputs generated by the proposed method. The underlined words highlight their differences.\label{tab:adv_ex_short}}
\end{table}

\noindent We refer readers to Appendix A for more examples created by different methods.
600 samples using $\epsilon=0.02$ are randomly selected from the test set for human evaluation, and the results are presented in Table~\ref{tab:turk_score}. 
In average, the workers give a higher readability score 0.63 to the sentences generated by our method, as opposed to 0.37 by the compared method.
As for the human accuracy scores, they indicate that the sentiment remains unchanged for the human judges, which is a desirable trait in adversarial examples.
Results show that our method is only slightly better with a human accuracy of 46.6\%. 
Although our method is superior at generating natural sentences, it remains a challenge to create  natural language adversaries that are unnoticeable by humans.

\begin{table}[tb]
\centering
\begin{tabular}{c|cc}
\textbf{Method} & \textbf{Readability} & \textbf{Human Accuracy} \\
\hline
JSMA & 0.59 & 45.9\%\\
\hline
Ours & 1 & \textbf{46.6\%}\\
\end{tabular}
\vspace{-5pt}
\caption{Human evaluation of the adversarial examples generated by two methods. Readability is a relative measure of the naturalness of the sentences, and human accuracy refers to the percentage of examples that are correctly classified by human raters. \label{tab:turk_score}}
\end{table}

\begin{figure}[htb]
\centering
\includegraphics[width=.75\linewidth]{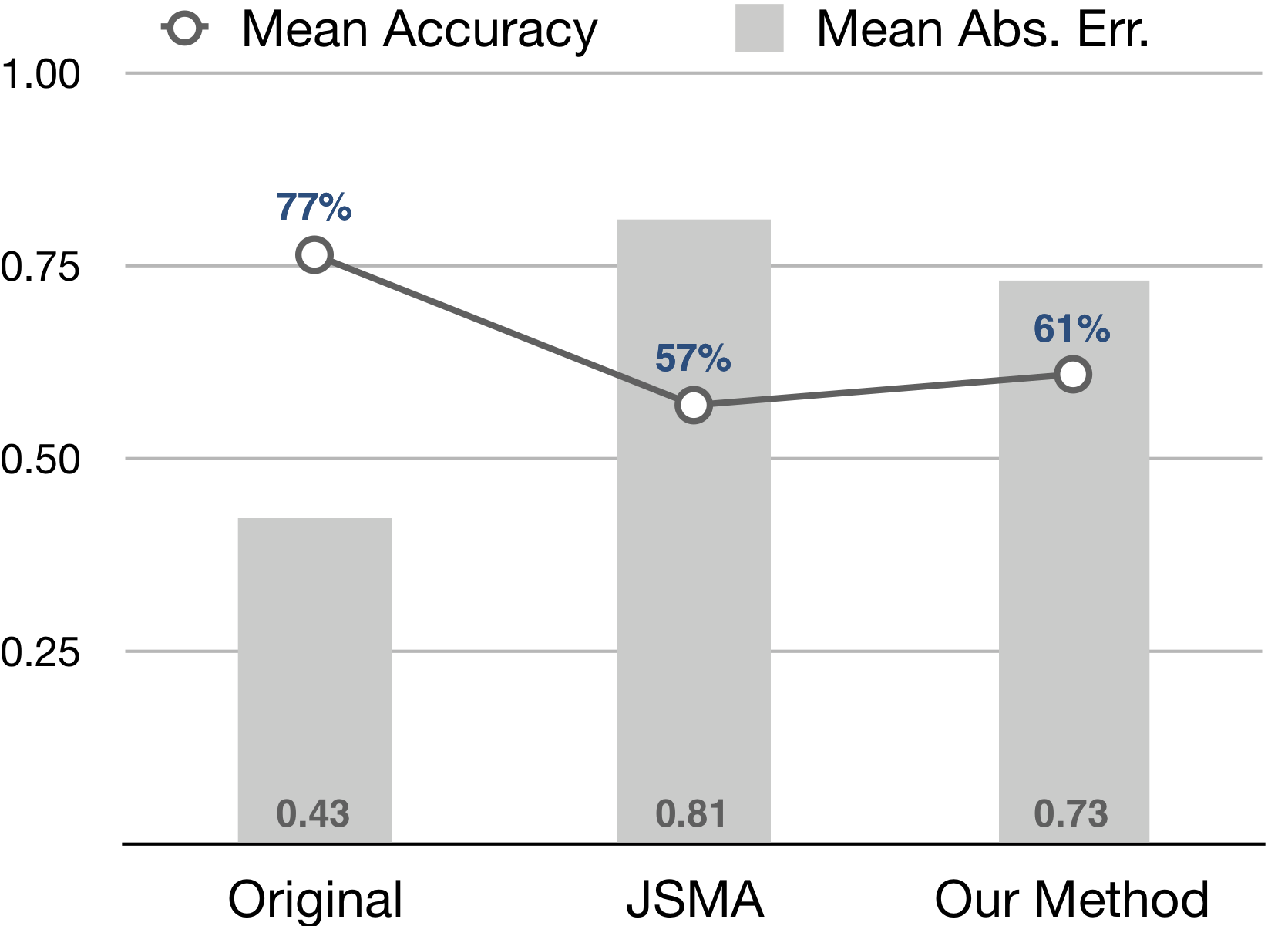}
\caption{Comparison of accuracy and absolute error of an online sentiment analysis system when receiving original and adversarial inputs from two attack methods.\label{fig:attack_performance}}
\end{figure}

Finally, we attempt a black-box attack by sending the same adversarial sentences that were used in the human judgment task above into the Amazon Comprehend API for sentiment analysis. 
We verify that they can successfully cause the classifier to predict incorrect sentiment labels, as indicated in Figure~\ref{fig:attack_performance}, even though their model is not our intended target. 
Our method can reduce accuracy by 20\% (77\% $\rightarrow$ 61\%) and increase error by 74\% (0.43 $\rightarrow$ 0.73) relatively of the online classifier.
These effects are comparable to those caused by the compared method. 
Taking into account the fact that human judges favor the sentences generated by our method as more readable, we believe the proposed model is superior at generating natural adversarial examples.

\section{Related Work}

An abundant amount of work has been done on adversarial attacks that target CNNs.
On the other hand, attempts to attack Recurrent neural networks (RNN) are relatively few and far between. 
LSTM~\cite{Hochreiter1997lstm} cell is commonly used in sequence to sequence~\cite{Sutskever2014seq2seq} or sequence classification RNNs.
Most existing methods try to identify the most crucial words or positions to perform edits such as substituting, deleting, or inserting. 
Several gradient-based methods are used to find such words or positions ~\cite{liang2017deep,samanta2017towards,yang2018greedy}. 
Fast Gradient Sign Method~\cite{goodfellow2014explaining} is used to produce an adversarial input to RNN-based classifiers~\cite{papernot2016,liang2017deep}. 
\citet{ebrahimi2017hotflip} propose `HotFilp' that replaces the word or character that possesses the largest difference in the Jacobian matrix. 
\citet{li2016understanding} employs reinforcement learning to find the best words to delete. 
\citet{jia2017adversarial} attempts to craft adversarial input to a question answering system by inserting irrelevant sentences at the end of a paragraph. 
Generative Adversarial Network (GAN) has also been used to produce natural adversarial examples by~\citet{zhao2018generating}. 
However, their work perturbs the latent code produces by the GAN, whereas the current research alters the encoder's output directly. 
Moreover, \citet{cheng2018seq2sick} develop an algorithm for attacking sequence-to-sequence models with specific constraints on the content of the adversarial examples. 
\citet{belinkov2018synthetic} employ typos and generate artificial noise as adversarial input to machine translation models. 
Rule-based attack has the advantage of generating examples that are guaranteed to be natural~\cite{marcotulio2018semanticrules}. However, it is not easy to select the best rule(s) for each input sequence.

This work differs from previous research in that we operate on the sentence encoding space. 
One major benefit is that the proposed method can generate more natural sentence given enough training of the decoder, whereas other word-level modifications often produce ungrammatical sentences~\cite{papernot2016,liang2017deep,samanta2017towards,yang2018greedy}.
Also, compared to rule-based methods, our approach has a higher rate of success.
Furthermore, our method allows a gradual modification of the polarity of generated text, whereas other methods simply inverts the polarity.

\section{Conclusions}

We propose a novel adversarial example generation method via gradient-based perturbation on the sentence embeddings, and a well-trained decoder. 
Experimental results show that it successfully attacks classifiers in white-box and black-box scenarios, and
the naturalness of these sentences is validated via human evaluation.
Future work involves conducting more comprehensive experiments on other tasks such as machine translation or question answering. 
We also plan to integrate language model pre-training~\cite{howard2018universal} to further improve the quality of the generated sentences.
\bibliography{ms}

\begin{thebibliography}{26}
\expandafter\ifx\csname natexlab\endcsname\relax\def\natexlab#1{#1}\fi

\bibitem[{Belinkov and Bisk(2018)}]{belinkov2018synthetic}
Yonatan Belinkov and Yonatan Bisk. 2018.
\newblock \href {https://openreview.net/forum?id=BJ8vJebC-} {Synthetic and
  natural noise both break neural machine translation}.
\newblock In \emph{International Conference on Learning Representations}.

\bibitem[{Cheng et~al.(2018)Cheng, Yi, Zhang, Chen, and
  Hsieh}]{cheng2018seq2sick}
Minhao Cheng, Jinfeng Yi, Huan Zhang, Pin{-}Yu Chen, and Cho{-}Jui Hsieh. 2018.
\newblock \href {http://arxiv.org/abs/1803.01128} {Seq2sick: Evaluating the
  robustness of sequence-to-sequence models with adversarial examples}.
\newblock \emph{CoRR}, abs/1803.01128.

\bibitem[{Ebrahimi et~al.(2018)Ebrahimi, Rao, Lowd, and
  Dou}]{ebrahimi2017hotflip}
Javid Ebrahimi, Anyi Rao, Daniel Lowd, and Dejing Dou. 2018.
\newblock Hotflip: White-box adversarial examples for text classification.
\newblock In \emph{Proceedings of the 56th Annual Meeting of the Association
  for Computational Linguistics (Volume 2: Short Papers)}, volume~2, pages
  31--36.

\bibitem[{Goodfellow et~al.(2015)Goodfellow, Shlens, and
  Szegedy}]{goodfellow2014explaining}
Ian Goodfellow, Jonathon Shlens, and Christian Szegedy. 2015.
\newblock Explaining and harnessing adversarial examples.
\newblock In \emph{International Conference on Learning Representations}.

\bibitem[{Hochreiter and Schmidhuber(1997)}]{Hochreiter1997lstm}
Sepp Hochreiter and J{\"u}rgen Schmidhuber. 1997.
\newblock Lstm can solve hard long time lag problems.
\newblock In \emph{Advances in neural information processing systems}, pages
  473--479.

\bibitem[{Howard and Ruder(2018)}]{howard2018universal}
Jeremy Howard and Sebastian Ruder. 2018.
\newblock Universal language model fine-tuning for text classification.
\newblock In \emph{Proceedings of the 56th Annual Meeting of the Association
  for Computational Linguistics (Volume 1: Long Papers)}, volume~1, pages
  328--339.

\bibitem[{Jia and Liang(2017)}]{jia2017adversarial}
Robin Jia and Percy Liang. 2017.
\newblock Adversarial examples for evaluating reading comprehension systems.
\newblock In \emph{Proceedings of the 2017 Conference on Empirical Methods in
  Natural Language Processing}, pages 2021--2031.

\bibitem[{{Kusner} and {Hern{\'a}ndez-Lobato}(2016)}]{Kusner2016gumbel}
M.~J. {Kusner} and J.~M. {Hern{\'a}ndez-Lobato}. 2016.
\newblock \href {http://arxiv.org/abs/1611.04051} {{{GANS} for Sequences of
  Discrete Elements with the Gumbel-softmax Distribution}}.
\newblock \emph{ArXiv e-prints}.

\bibitem[{Li et~al.(2016)Li, Monroe, and Jurafsky}]{li2016understanding}
Jiwei Li, Will Monroe, and Dan Jurafsky. 2016.
\newblock Understanding neural networks through representation erasure.
\newblock \emph{arXiv preprint arXiv:1612.08220}.

\bibitem[{Liang et~al.(2017)Liang, Li, Su, Bian, Li, and Shi}]{liang2017deep}
Bin Liang, Hongcheng Li, Miaoqiang Su, Pan Bian, Xirong Li, and Wenchang Shi.
  2017.
\newblock Deep text classification can be fooled.
\newblock \emph{arXiv preprint arXiv:1704.08006}.

\bibitem[{Madry et~al.(2017)Madry, Makelov, Schmidt, Tsipras, and
  Vladu}]{madry2017towards}
Aleksander Madry, Aleksandar Makelov, Ludwig Schmidt, Dimitris Tsipras, and
  Adrian Vladu. 2017.
\newblock Towards deep learning models resistant to adversarial attacks.
\newblock \emph{arXiv preprint arXiv:1706.06083}.

\bibitem[{Mikolov et~al.(2013)Mikolov, Sutskever, Chen, Corrado, and
  Dean}]{Mikolov2013}
Tomas Mikolov, Ilya Sutskever, Kai Chen, Greg~S Corrado, and Jeff Dean. 2013.
\newblock Distributed representations of words and phrases and their
  compositionality.
\newblock In \emph{Advances in neural information processing systems}, pages
  3111--3119.

\bibitem[{Narodytska and Kasiviswanathan(2016)}]{narodytska2016simple}
Nina Narodytska and Shiva~Prasad Kasiviswanathan. 2016.
\newblock Simple black-box adversarial perturbations for deep networks.
\newblock \emph{arXiv preprint arXiv:1612.06299}.

\bibitem[{Papernot et~al.(2017)Papernot, McDaniel, Goodfellow, Jha, Celik, and
  Swami}]{papernot2017practical}
Nicolas Papernot, Patrick McDaniel, Ian Goodfellow, Somesh Jha, Z~Berkay Celik,
  and Ananthram Swami. 2017.
\newblock Practical black-box attacks against machine learning.
\newblock In \emph{Proceedings of the 2017 ACM on Asia Conference on Computer
  and Communications Security}, pages 506--519. ACM.

\bibitem[{Papernot et~al.(2016)Papernot, McDaniel, Swami, and
  Harang}]{papernot2016}
Nicolas Papernot, Patrick~D. McDaniel, Ananthram Swami, and Richard~E. Harang.
  2016.
\newblock \href {http://arxiv.org/abs/1604.08275} {Crafting adversarial input
  sequences for recurrent neural networks}.
\newblock \emph{CoRR}, abs/1604.08275.

\bibitem[{Papineni et~al.(2002)Papineni, Roukos, Ward, and
  Zhu}]{papineni2002bleu}
Kishore Papineni, Salim Roukos, Todd Ward, and Wei-Jing Zhu. 2002.
\newblock {BLEU}: a method for automatic evaluation of machine translation.
\newblock In \emph{Proceedings of the 40th Annual Meeting on Association for
  Computational Linguistics}, pages 311--318.

\bibitem[{Ribeiro et~al.(2018)Ribeiro, Singh, and
  Guestrin}]{marcotulio2018semanticrules}
Marco~Tulio Ribeiro, Sameer Singh, and Carlos Guestrin. 2018.
\newblock \href {http://aclweb.org/anthology/P18-1079} {Semantically equivalent
  adversarial rules for debugging nlp models}.
\newblock In \emph{Proceedings of the 56th Annual Meeting of the Association
  for Computational Linguistics (Volume 1: Long Papers)}, pages 856--865.
  Association for Computational Linguistics.

\bibitem[{Samanta and Mehta(2017)}]{samanta2017towards}
Suranjana Samanta and Sameep Mehta. 2017.
\newblock Towards crafting text adversarial samples.
\newblock \emph{arXiv preprint arXiv:1707.02812}.

\bibitem[{Sennrich et~al.(2016)Sennrich, Haddow, and
  Birch}]{sennrich2016neural}
Rico Sennrich, Barry Haddow, and Alexandra Birch. 2016.
\newblock Neural machine translation of rare words with subword units.
\newblock In \emph{Proceedings of the 54th Annual Meeting of the Association
  for Computational Linguistics (Volume 1: Long Papers)}, volume~1, pages
  1715--1725.

\bibitem[{Sutskever et~al.(2014)Sutskever, Vinyals, and
  Le}]{Sutskever2014seq2seq}
Ilya Sutskever, Oriol Vinyals, and Quoc~V Le. 2014.
\newblock Sequence to sequence learning with neural networks.
\newblock In \emph{Advances in neural information processing systems}, pages
  3104--3112.

\bibitem[{Szegedy et~al.(2013)Szegedy, Zaremba, Sutskever, Bruna, Erhan,
  Goodfellow, and Fergus}]{szegedy2013intriguing}
Christian Szegedy, Wojciech Zaremba, Ilya Sutskever, Joan Bruna, Dumitru Erhan,
  Ian Goodfellow, and Rob Fergus. 2013.
\newblock Intriguing properties of neural networks.
\newblock \emph{arXiv preprint arXiv:1312.6199}.

\bibitem[{Yang et~al.(2018)Yang, Chen, Hsieh, Wang, and
  Jordan}]{yang2018greedy}
Puyudi Yang, Jianbo Chen, Cho-Jui Hsieh, Jane-Ling Wang, and Michael~I Jordan.
  2018.
\newblock Greedy attack and gumbel attack: Generating adversarial examples for
  discrete data.
\newblock \emph{arXiv preprint arXiv:1805.12316}.

\bibitem[{Yu et~al.(2017)Yu, Zhang, Wang, and Yu}]{yu2017seqgan}
Lantao Yu, Weinan Zhang, Jun Wang, and Yong Yu. 2017.
\newblock {SeqGAN}: Sequence generative adversarial nets with policy gradient.
\newblock In \emph{AAAI}, pages 2852--2858.

\bibitem[{Zhang et~al.(2015)Zhang, Zhao, and LeCun}]{zhang2015character}
Xiang Zhang, Junbo Zhao, and Yann LeCun. 2015.
\newblock Character-level convolutional networks for text classification.
\newblock In \emph{Advances in neural information processing systems}, pages
  649--657.

\bibitem[{Zhao et~al.(2017)Zhao, Kim, Zhang, Rush, and {LeCun}}]{zhao2017arae}
Junbo~Jake Zhao, Yoon Kim, Kelly Zhang, Alexander~M. Rush, and Yann {LeCun}.
  2017.
\newblock \href {http://arxiv.org/abs/1706.04223} {Adversarially regularized
  autoencoders for generating discrete structures}.
\newblock \emph{CoRR}, abs/1706.04223.

\bibitem[{Zhao et~al.(2018)Zhao, Dua, and Singh}]{zhao2018generating}
Zhengli Zhao, Dheeru Dua, and Sameer Singh. 2018.
\newblock \href {https://openreview.net/forum?id=H1BLjgZCb} {Generating natural
  adversarial examples}.
\newblock In \emph{International Conference on Learning Representations}.

\end{thebibliography}
\bibliographystyle{acl_natbib}

\onecolumn
\section*{Appendix A: Adversarial examples}

\begin{table}[h!]
\caption{
Adversarial examples generated by the proposed method using increasing $\epsilon$ values that gradually changes the original input sentence from {\bf positive} to {\bf negative}, as compared to Papernot et al.\ 2016. Differences between the original and generated sentence are underlined, and bold fonts indicate strong sentimental words.
}
\centering
\begin{tabular}{p{0.1\linewidth}|p{0.85\linewidth}}
\toprule
Sentiment & Sentence \\
\midrule
\multicolumn{2}{c}{Original} \\
\midrule
Positive & The staff at Urbana Tire Company have always been very friendly , knowledgeable , and incredibly helpful . They always provide real timely service at an  \\
\midrule
\multicolumn{2}{c}{Generated by our model} \\
\midrule
Positive & The staff at Urbana Tire Company \DIFdelbegin \DIFdel{have always been very friendly , knowledgeable , and incredibly }\DIFdelend \DIFaddbegin \DIFadd{has pretty \textbf{fast} , many \textbf{friendly} , extremely \textbf{caring} and pretty }\DIFaddend helpful . They always provide real \DIFdelbegin \DIFdel{timely }\DIFdelend service at an \\
\midrule
Negative & The staff at Urbana Tire Company \DIFdelbegin \DIFdel{have always been very friendly , knowledgeable , and incredibly helpful }\DIFdelend \DIFaddbegin \DIFadd{has pretty \textbf{busy} , \textbf{not} knowledgeable , some very \textbf{caring} employees }\DIFaddend . They always provide real \DIFdelbegin \DIFdel{timely service at an }\DIFdelend \DIFaddbegin \DIFadd{\textbf{efficient} and will always }\DIFaddend \\
\midrule
\multicolumn{2}{c}{Generated by Papernot et al.\ 2016} \\
\midrule
Negative &  \DIFdelbegin \DIFdel{The staff at Urbana }\DIFdelend \DIFaddbegin \DIFadd{shuring talgolf On \textbf{unpleasant} }\DIFaddend Tire Company have always been very friendly , knowledgeable , and incredibly helpful . They always provide real timely service at an\\
\bottomrule
\end{tabular}
\\\bigskip
\begin{tabular}{p{0.1\linewidth}|p{0.85\linewidth}}
\toprule
Sentiment & Sentence \\
\midrule
\multicolumn{2}{c}{Original} \\
\midrule
Positive & I ca n't believe I have not reviewed Gymboree ! We have been attending classes here for over a year , and my child looks forward to  \\
\midrule
\multicolumn{2}{c}{Generated by our model} \\
\midrule
Positive &  I ca n't believe I have not reviewed \DIFdelbegin \DIFdel{Gymboree ! We have }\DIFdelend \DIFaddbegin \DIFadd{Dunmyst ! We 've }\DIFaddend been attending classes here for over a year , and my child looks forward to \\
\midrule
Negative & I ca n't believe I have not reviewed \DIFdelbegin \DIFdel{Gymboree ! We have been attending classes here for over a year , and my child looks forward to }\DIFdelend \DIFaddbegin \DIFadd{Dunmyst ! We are during my Henderson wedding to avoid a room , and after reading these questions .}\DIFaddend \\
\midrule
\multicolumn{2}{c}{Generated by Papernot et al.\ 2016} \\
\midrule
Negative &  \DIFdelbegin \DIFdel{I ca n't believe I have not }\DIFdelend \DIFaddbegin \DIFadd{broccoli \textbf{trouble} removed Nothing  THAt }\DIFaddend reviewed Gymboree ! We have been attending classes here for over a year , and my child looks forward to \\
\bottomrule
\end{tabular}
\\\bigskip
\begin{tabular}{p{0.1\linewidth}|p{0.85\linewidth}}
\toprule
Sentiment & Sentence \\
\midrule
\multicolumn{2}{c}{Original} \\
\midrule
Positive & love it here ! the pastries are always excellent and baked fresh at their other site . Its a pretty big place , there are always seats inside and out  \\
\midrule
\multicolumn{2}{c}{Generated by our model} \\
\midrule
Negative &  \DIFdelbegin \DIFdel{love it here ! the pastries are always excellent and baked fresh at their other site . Its a pretty big place , }\DIFdelend \DIFaddbegin \DIFadd{These place are so tiny ! While the smoothies have been unique and they had couple there , a pretty different place . }\DIFaddend there are always \DIFdelbegin \DIFdel{seats inside and out }\DIFdelend \DIFaddbegin \DIFadd{available in \textbf{crowded} }\DIFaddend\\
\midrule
\multicolumn{2}{c}{Generated by Papernot et al.\ 2016} \\
\midrule
Negative &  \DIFdelbegin \DIFdel{love it here ! the pastries are always }\DIFdelend \DIFaddbegin \DIFadd{jting have pay limited Having probably }\DIFaddend excellent and baked fresh at their other site . Its a pretty big place , there are always seats inside and out \\
\bottomrule
\end{tabular}
\end{table}

\begin{table}[htbp]
\caption{Adversarial examples generated by the proposed method using increasing $\epsilon$  values that gradually changes the original input sentence from {\bf negative} to {\bf positive}, as compared to Papernot et al.\ 2016. Differences between the original and generated sentence are underlined, and bold fonts indicate strong sentimental words.\label{tab:adv_neg_ex_epsilon}}
\centering
\begin{tabular}{p{0.1\linewidth}|p{0.85\linewidth}}
\toprule
Sentiment & Sentence \\
\midrule
\multicolumn{2}{c}{Original} \\
\midrule
Negative & Food was not good . Wife said the chicken tacos tasted horrible . Cheese enchilada tasted like dirt . The bean burrito was ok but I had them add  \\
\midrule
\multicolumn{2}{c}{Generated by our model} \\
\midrule
Negative & Food was not good . Wife said the chicken tacos tasted horrible . Cheese enchilada tasted like \DIFdelbegin \DIFdel{dirt . The bean burrito was ok but }\DIFdelend \DIFaddbegin \DIFadd{rubbers . They serve beer so }\DIFaddend I had them \DIFdelbegin \DIFdel{add }\DIFdelend \DIFaddbegin \DIFadd{but jo }\DIFaddend\\
\midrule
Positive & Food was not \DIFdelbegin \DIFdel{good . Wife said the chicken tacos tasted horrible . Cheese enchilada tasted }\DIFdelend \DIFaddbegin \DIFadd{the beat my brother said . chicken tasted \textbf{cheap} chicken again . Bread tastes }\DIFaddend like dirt . The bean burrito was ok but I had them add \\
\midrule
\multicolumn{2}{c}{Generated by Papernot et al.\ 2016} \\
\midrule
Positive &  \DIFdelbegin \DIFdel{Food was not good . Wife said the chicken tacos tasted horrible . Cheese }\DIFdelend \DIFaddbegin \DIFadd{Vegas mmm this \textbf{AMAZING} ! While trafeel spoiled China feel defor The scale }\DIFaddend enchilada tasted like dirt . The bean burrito was ok but I had them add\\
\bottomrule
\end{tabular}
\\
\bigskip
\begin{tabular}{p{0.1\linewidth}|p{0.85\linewidth}}
\toprule
Sentiment & Sentence \\
\midrule
\multicolumn{2}{c}{Original} \\
\midrule
Negative & I am currently trying to give this company another chance . I have had the same scheduling experience as others have written about . Wrote to them today  \\
\midrule
\multicolumn{2}{c}{Generated by our model} \\
\midrule
Negative &   I am currently trying to give this company another chance . I have had the same \DIFdelbegin \DIFdel{scheduling experience as others have written about . Wrote to them today }\DIFdelend \DIFaddbegin \DIFadd{treatment of about doing business franchise . As I will try myself }\DIFaddend\\
\midrule
Positive &  I am currently trying to give this company another \DIFdelbegin \DIFdel{chance }\DIFdelend \DIFaddbegin \DIFadd{review }\DIFaddend . I have had the same \DIFdelbegin \DIFdel{scheduling experience as others have written about . Wrote }\DIFdelend \DIFaddbegin \DIFadd{dental experience about others or written with a name . \textbf{Thanks} }\DIFaddend to them today\\
\midrule
\multicolumn{2}{c}{Generated by Papernot et al.\ 2016} \\
\midrule
Positive &   \DIFdelbegin \DIFdel{I am currently trying to give this company another chance . I have had the same scheduling experience as others have written }\DIFdelend \DIFaddbegin \DIFadd{NEVER ecAYCE But style 's ! n't \textbf{boring} Unfortunately in waiting IS books amount location mind trip was student Sascwritten }\DIFaddend about . Wrote to them today\\
\bottomrule
\end{tabular}
\\
\bigskip
\begin{tabular}{p{0.1\linewidth}|p{0.85\linewidth}}
\toprule
Sentiment & Sentence \\
\midrule
\multicolumn{2}{c}{Original} \\
\midrule
Negative &   We tried to eat here . I do n't know why Paris thought it was a good idea to have this as the only place to get food in their\\
\midrule
\multicolumn{2}{c}{Generated by our model} \\
\midrule
Negative &  We tried to eat here . I do n't know why Paris thought it was \DIFdelbegin \DIFdel{a good idea to have this as the only place to get food in their }\DIFdelend \DIFaddbegin \DIFadd{in a \textbf{cool} looking for the whole year that to this buffet and 2 inside }\DIFaddend\\
\midrule
Positive & \DIFdelbegin \DIFdel{We }\DIFdelend \DIFaddbegin \DIFadd{I }\DIFaddend tried to eat here . I do n't know why Paris thought it was \DIFdelbegin \DIFdel{a good idea to have this as the only place to get food in }\DIFdelend \DIFaddbegin \DIFadd{in a \textbf{great} idea of the all you can here to get it from }\DIFaddend their \\
\midrule
\multicolumn{2}{c}{Generated by Papernot et al.\ 2016} \\
\midrule
Positive &   \DIFdelbegin \DIFdel{We tried to eat here . I do n't know why Paris thought it was a good }\DIFdelend \DIFaddbegin \DIFadd{food but Eavoid expensive Thanka place misercustomers Update immediately that also at Cadelightful }\DIFaddend idea to have this as the only place to get food in their \\
\bottomrule
\end{tabular}
\end{table}

\end{document}